\documentclass[aps,prd,twocolumn,superscriptaddress,showpacs]{revtex4}
\usepackage{graphicx}
\usepackage{amsmath,amssymb,latexsym}
\usepackage{bm}
\usepackage{url}

\begin{document}

\title{Cosmogenic photons as a test of ultra-high energy cosmic ray
  composition}

\author{Dan Hooper} \affiliation{Center for Particle Astrophysics,
  Fermi National Accelerator Laboratory, Batavia, IL 60510, USA}
\affiliation{Department of Astronomy and Astrophysics, The University
  of Chicago, Chicago, IL 60637, USA} 
\author{Andrew M. Taylor}
\affiliation{ISDC, Chemin d'Ecogia 16, Versoix, CH-1290, Switzerland}
\author{Subir Sarkar} \affiliation{Rudolf Peierls Centre for
  Theoretical Physics, University of Oxford, Oxford OX1 3NP, UK}


\begin{abstract}

  Although recent measurements of the shower profiles of ultra-high energy
  cosmic rays suggest that they are largely initiated by heavy
  nuclei, such conclusions rely on hadronic interaction models
  which have large uncertainties. We investigate an alternative test
  of cosmic ray composition which is based on the observation of
  ultra-high energy photons produced  through cosmic ray interactions with diffuse 
  low energy photon backgrounds during intergalactic propagation. We
  show that if the ultra-high energy cosmic rays are
  dominated by heavy nuclei, the flux of these photons is
  suppressed by approximately an order of magnitude relative to the
  proton-dominated case. Future observations by the Pierre Auger
  Observatory may be able to use this observable to constrain the
  composition of the primaries, thus providing an important cross-check of
  hadronic interaction models.
\end{abstract}

\pacs{96.50.S-,98.70.Sa,13.85.Tp; FERMILAB-PUB-10-219-A}

\maketitle

Despite considerable experimental and theoretical effort, the chemical
composition of the ultra-high energy cosmic rays (UHECRs) remains
ambiguous. Recent measurements of air shower profiles by the Pierre
Auger Observatory (PAO) suggest that UHECRs are increasingly dominated
by heavy nuclei at energies above $10^{18.5}$~eV \cite{xmax,heavy}. Uncertainties at such high energies in the hadronic interaction models
used to interpret the data \cite{hadronic}, however, can undermine this
conclusion \cite{uncertain}. In this paper, we discuss a complementary observation
that, without relying on hadronic interaction models, can be used to
constrain the chemical composition of the UHECRs.

Protons with energy above $\sim$$10^{19.6}$ eV (the ``GZK cutoff'' energy)
interact efficiently with the cosmic microwave (and infrared)
background, producing charged and neutral pions \cite{gzk} whose
decays yield potentially observable fluxes of UHE neutrinos and
photons. The detection of these ``cosmogenic neutrinos''~\cite{cosmogenic} is a key target
for present \cite{limits} and planned high energy neutrino telescopes
\cite{interest}. The PAO has placed stringent limits on the fraction
of UHECRs that are photons \cite{augersdphotonfractionlimit} and is
expected to ultimately reach the level of sensitivity required to detect the
cosmogenic photon flux \cite{Risse:2007sd}.

If UHECRs are largely heavy or intermediate mass nuclei, however, they
will interact with radiation backgrounds primarily through
photo-disintegration, breaking up into lighter nuclei and nucleons. As
these nucleons are often below the energy threshold for
pion-production, fewer UHE neutrinos and photons are produced. This
leads to significant suppression of the cosmogenic neutrino flux
\cite{neutrinos}. We describe here how a heavy chemical composition of
the UHECR spectrum also leads to suppression of the cosmogenic photon flux. Thus as
the PAO's sensitivity to UHE photons increases, this will provide a
new probe of the composition of UHECRs.
 
Following our previous work \cite{previous}, we simulate the
intergalactic propagation of UHECRs by an analytically validated Monte
Carlo method, including the effects of photo-pion and pair production
as well as photo-disintegration (for related work see
Ref.~\cite{other}). We assume that the UHECR sources are homogeneously
distributed and that they produce protons or nuclei with a power-law
spectrum up to a maximum energy, above which the flux is exponentially
suppressed: ${\rm d}N/{\rm d}E \propto E^{-\alpha} \exp(-E/E_{\rm
  max,Z})$. To maintain consistency with our previous work, we express
the maximum energy in terms of the quantity $E_{\rm max, Z} \equiv
E_{\rm max} \times (26/Z)$, where $Z$ is the electric charge of the
cosmic ray nucleus.

If the UHECRs are all protons, a good fit to the observed cosmic ray
spectrum above $10^{19}$ eV can be found for an injected spectrum with
a spectral index in the range $\alpha \approx 1.6-2.4$, and $E_{\rm
  max} \sim (1-5) \times 10^{21}$ eV; a similar range of spectral
indices can also provide reasonable fits for heavy or intermediate
mass UHECRs \cite{spectrum}. Henceforth we set $\alpha=2.0$, although
our results depend only weakly on the precise value
\cite{alphaprecise}. For iron, silicon, or nitrogen nuclei, we find
that the observed spectrum requires $E_{\rm max} \gtrsim 10^{20}$,
$10^{20.5}$, or $10^{21}$ eV, respectively. We do not consider values
of $E_{\rm max}$ greater than $10^{22}$~eV since there is no
plausible astrophysical source which can even contain such high energy
particles \cite{Hillas:1985is}.

Our Monte Carlo code tracks the propagation of each individual UHE
nucleus, nucleon, photon, and electron down to an energy of
$10^{18}$~eV. As they propagate, UHE photons produce $e^- e^+$ pairs
through interactions with the cosmic radio (and microwave) background 
at a rate given by
\begin{eqnarray}
  R (E_{\gamma}) = \frac{2 m^2_e}{E^2_\gamma} \int \frac{1}{\epsilon^2}
  \frac{{\rm d}n}{{\rm d}\epsilon} {\rm d}\epsilon 
  \int_0^{E_\gamma \epsilon/m_e} \epsilon^\prime  \sigma_{\gamma\gamma} 
  (E_\gamma, \epsilon^\prime) {\rm d} \epsilon^\prime, 
\end{eqnarray}
where $E_{\gamma}$ is the energy of the propagating photon, $\epsilon$
is the energy of the background photon, ${\rm d}n/{\rm d}\epsilon$
describes the background photon distribution, and
$\sigma_{\gamma\gamma} (E_\gamma, \epsilon)$ is the cross-section for
pair production. At energies above $10^{18}$~eV, the interaction
length of a photon is comparable to or shorter than that of UHE
protons and nuclei, {\em viz.} $\sim 1-10$ Mpc.

In each collision, the incoming photon transfers a significant
fraction of its energy to an outgoing electron or positron (a plot
showing this quantity for different center-of-mass energies is shown in, {\em e.g.},
Ref.~\cite{Taylor:2009iw}). For a $10^{19}$ eV ($10^{20}$ eV) photon
scattering off of a $10^{-6}$ eV radio photon, for example, more than
90\% (97\%) of the energy is transferred to the highly boosted
outgoing $e^-$/$e^+$.









UHE electrons and positrons produced in this manner can subsequently
regenerate an UHE photon through inverse Compton scattering with
CMB photons at a rate given by
\begin{eqnarray}
  R (E_e) = \frac{2 m^2_e}{E^2_e} \int \frac{1}{\epsilon^2} 
  \frac{{\rm d}n}{{\rm d}\epsilon} {\rm d} \epsilon 
  \int_0^{4E_e\epsilon/m_e} \epsilon^\prime \sigma_{e \gamma} 
  (E_e, \epsilon^\prime){\rm d}\epsilon^\prime.
\end{eqnarray}
Each collision transfers the bulk of the initial particle energy into
the photon. We follow the development of the resulting electromagnetic
cascade following the technique described in Ref.~\cite{Taylor:2008jz}
(see also Ref.~\cite{Eungwanichayapant:2009bi}).

Electrons and positrons can also lose energy through synchrotron
radiation in magnetic fields. Whether typical UHE electrons lose a
substantial fraction of their energy before inverse Compton scattering
depends on the relative energy densities of the extragalactic magnetic
field and the cosmic radio background. Competing with this effect is
the fact that UHE nuclei and protons will also be deflected by
magnetic fields, increasing their energy losses during
propagation. Taken together, we find that the presence of nano-Gauss
scale extragalactic magnetic fields increases slightly the
resulting fraction of UHECRs that are photons at energies $\sim 10^{18}$~eV,
and decreases the photon fraction at energies $> 10^{19}$~eV.

For the cosmic radio background we adopt the two extreme possibilities.
 The first is the estimate from observations given in Ref.~\cite{Clark:1970} 
which may well be contaminated by foreground
synchrotron emission from cosmic ray electrons in the galactic halo;
hence, following Ref.~\cite{Keshet:2004dr}, we consider this to represent an upper limit. We also present results for the case in which only the
radio component of the cosmic microwave background contributes,
representing a lower limit. We consider two specific extragalactic
magnetic field strengths, ranging from the observational upper
limit of $\sim$$10^{-9}$ G to (negligibly) weak values of
$3 \times 10^{-12}$~G~\cite{Taylor:2008jz}. These field strengths 
bound the range of possible effects that extragalactic magnetic fields
may have on the results.

\begin{figure*}[!t]
\begin{center}
\includegraphics[angle=0,width=0.325\linewidth,type=pdf,ext=.pdf,read=.pdf]{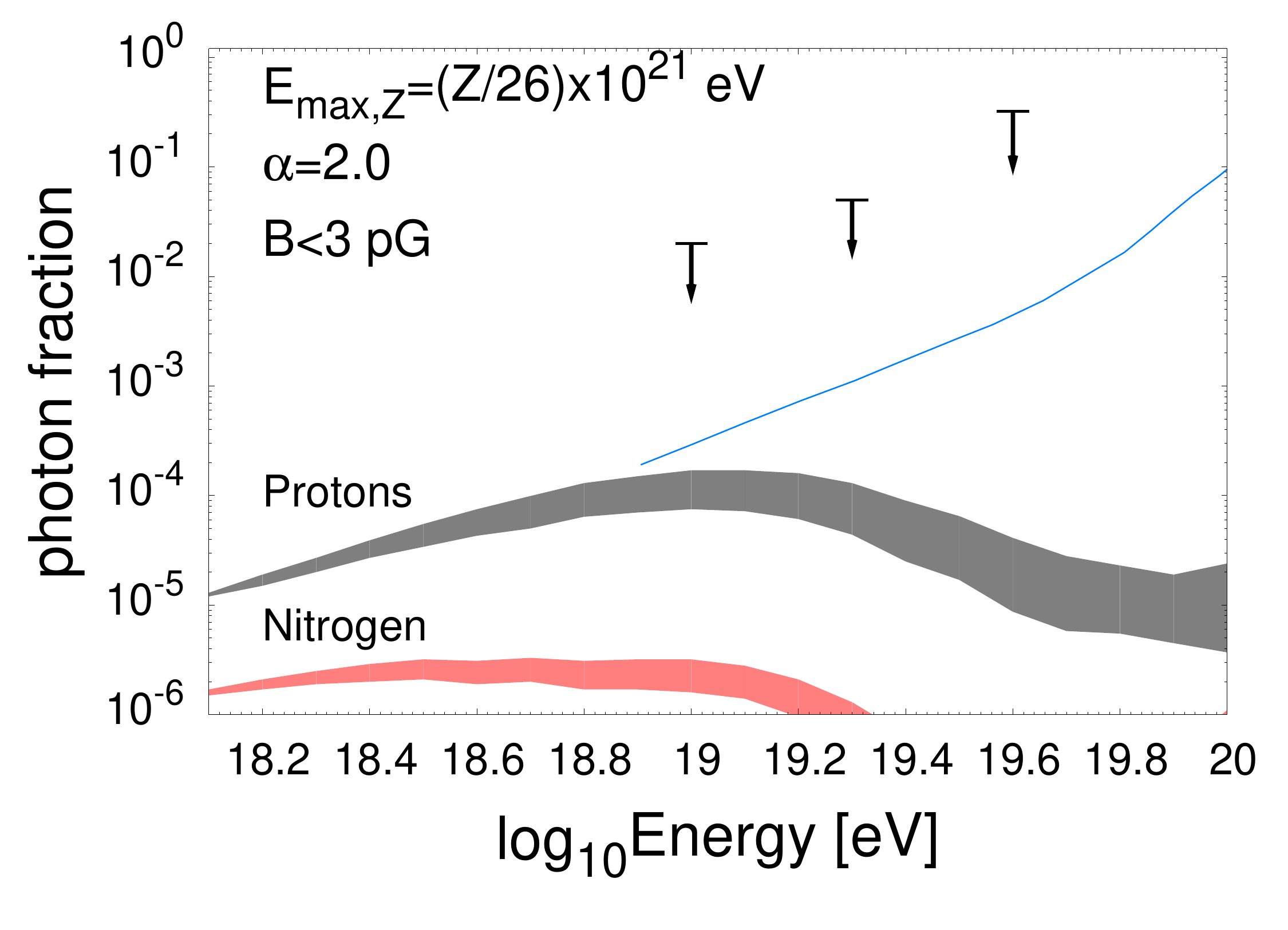}
\includegraphics[angle=0,width=0.325\linewidth,type=pdf,ext=.pdf,read=.pdf]{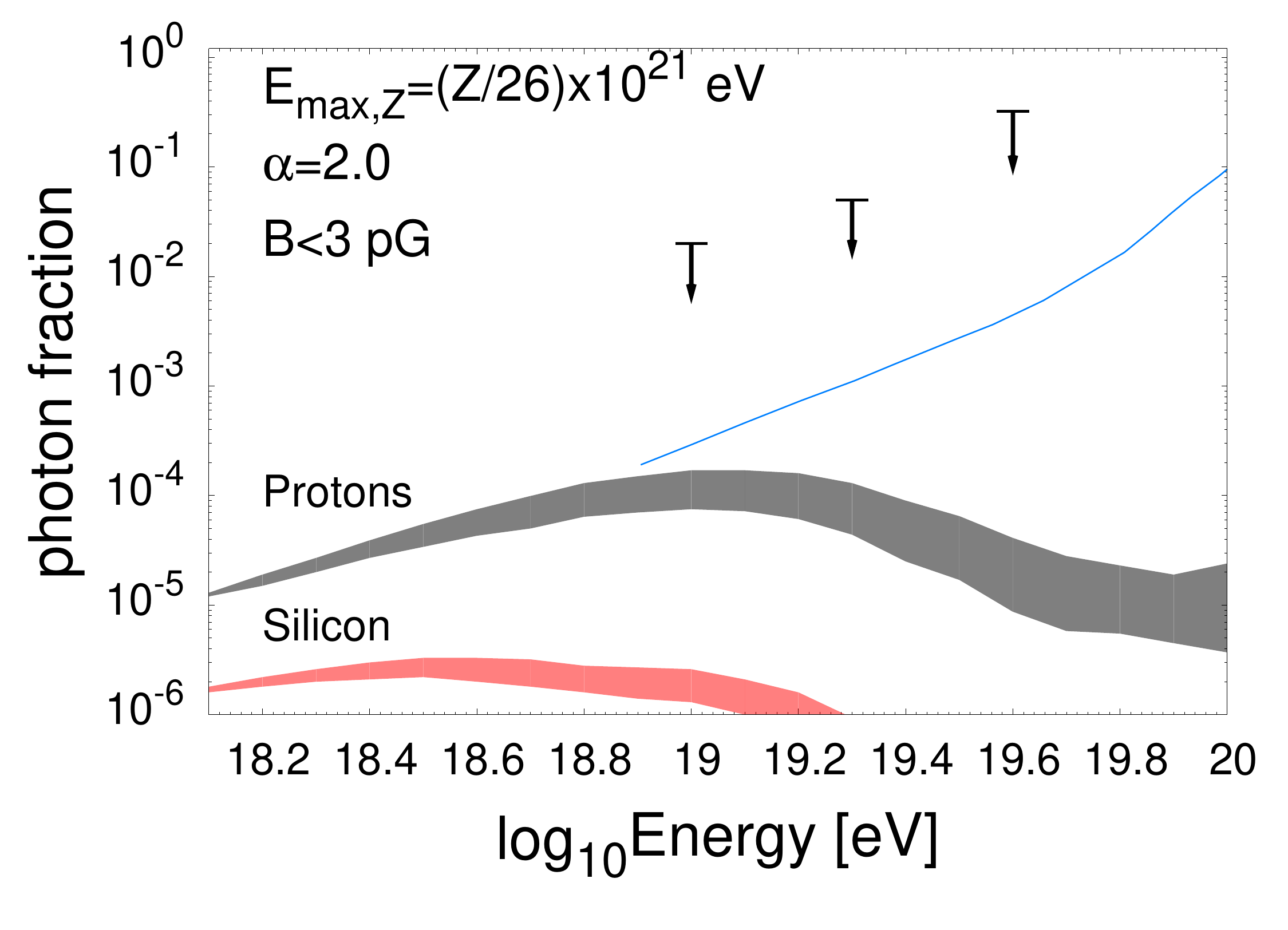}
\includegraphics[angle=0,width=0.325\linewidth,type=pdf,ext=.pdf,read=.pdf]{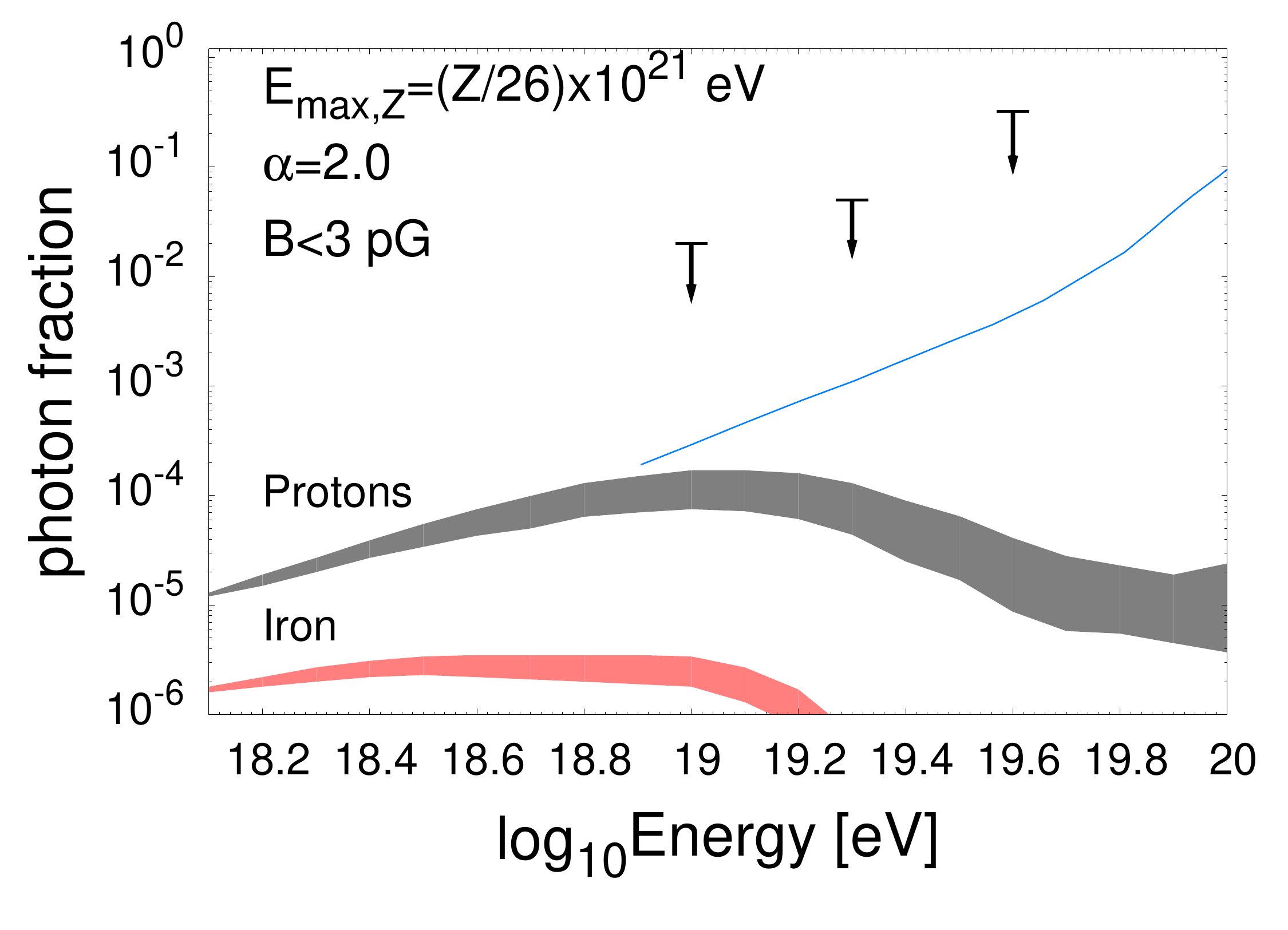}
\includegraphics[angle=0,width=0.325\linewidth,type=pdf,ext=.pdf,read=.pdf]{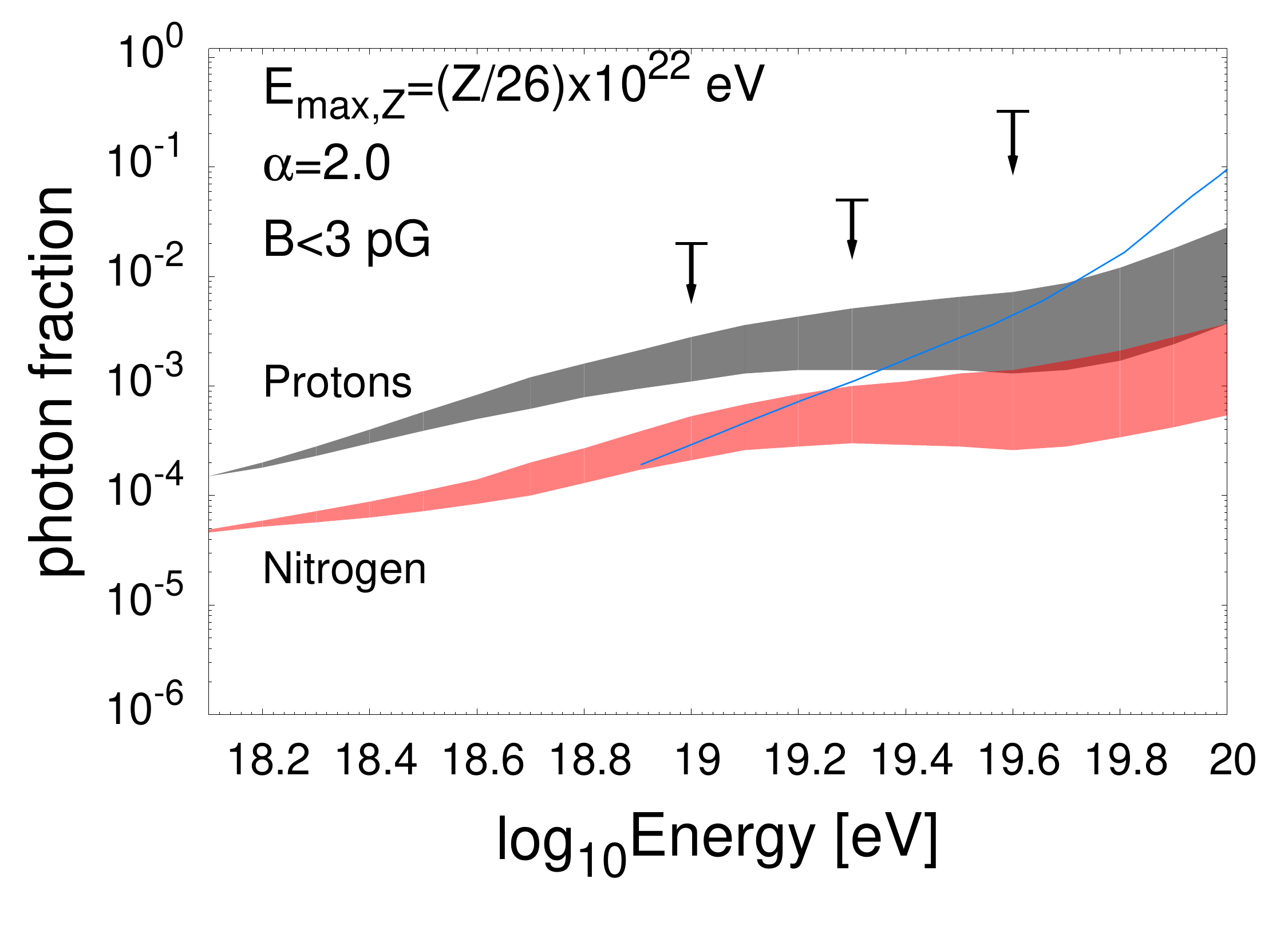}
\includegraphics[angle=0,width=0.325\linewidth,type=pdf,ext=.pdf,read=.pdf]{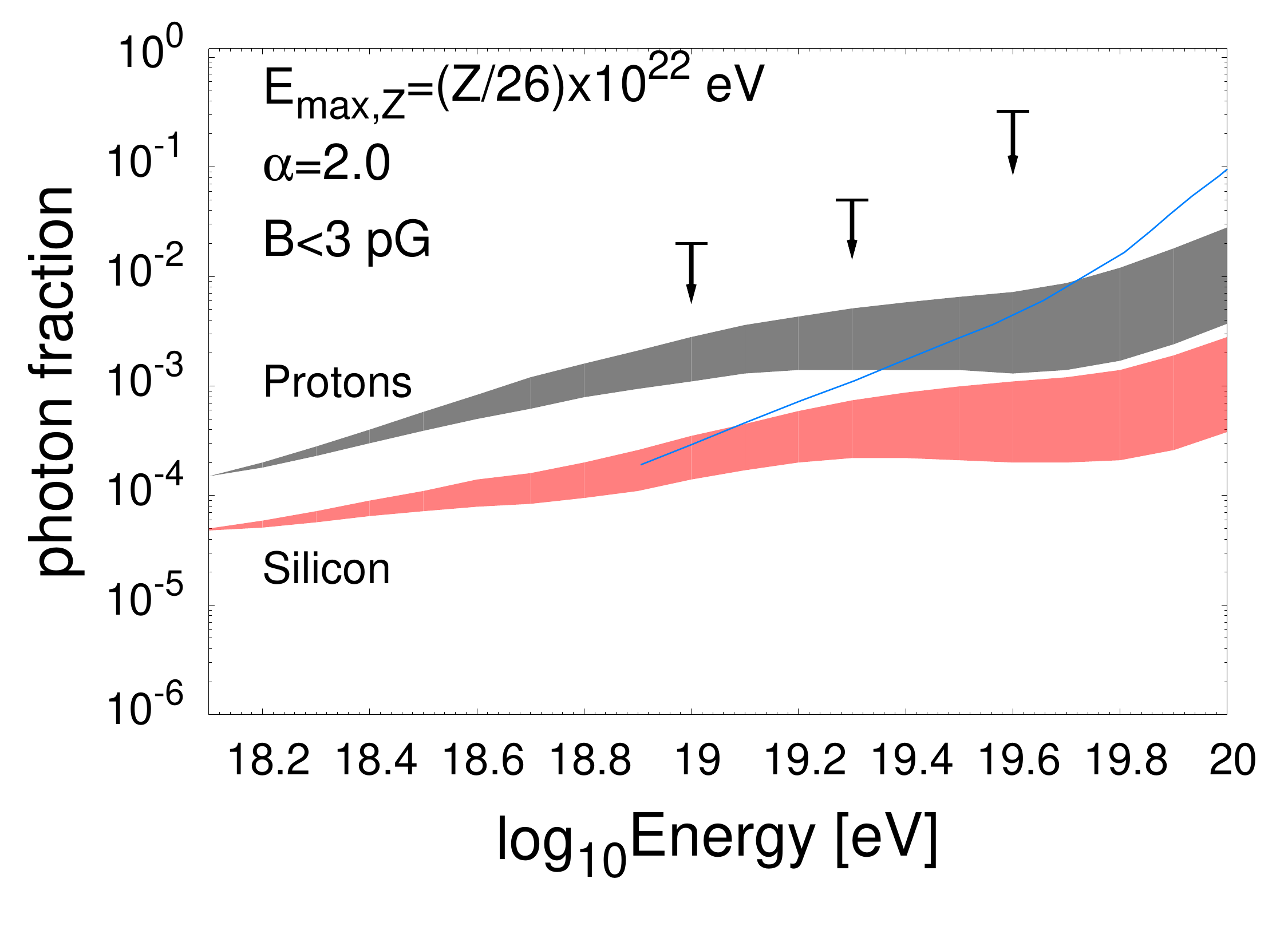}
\includegraphics[angle=0,width=0.325\linewidth,type=pdf,ext=.pdf,read=.pdf]{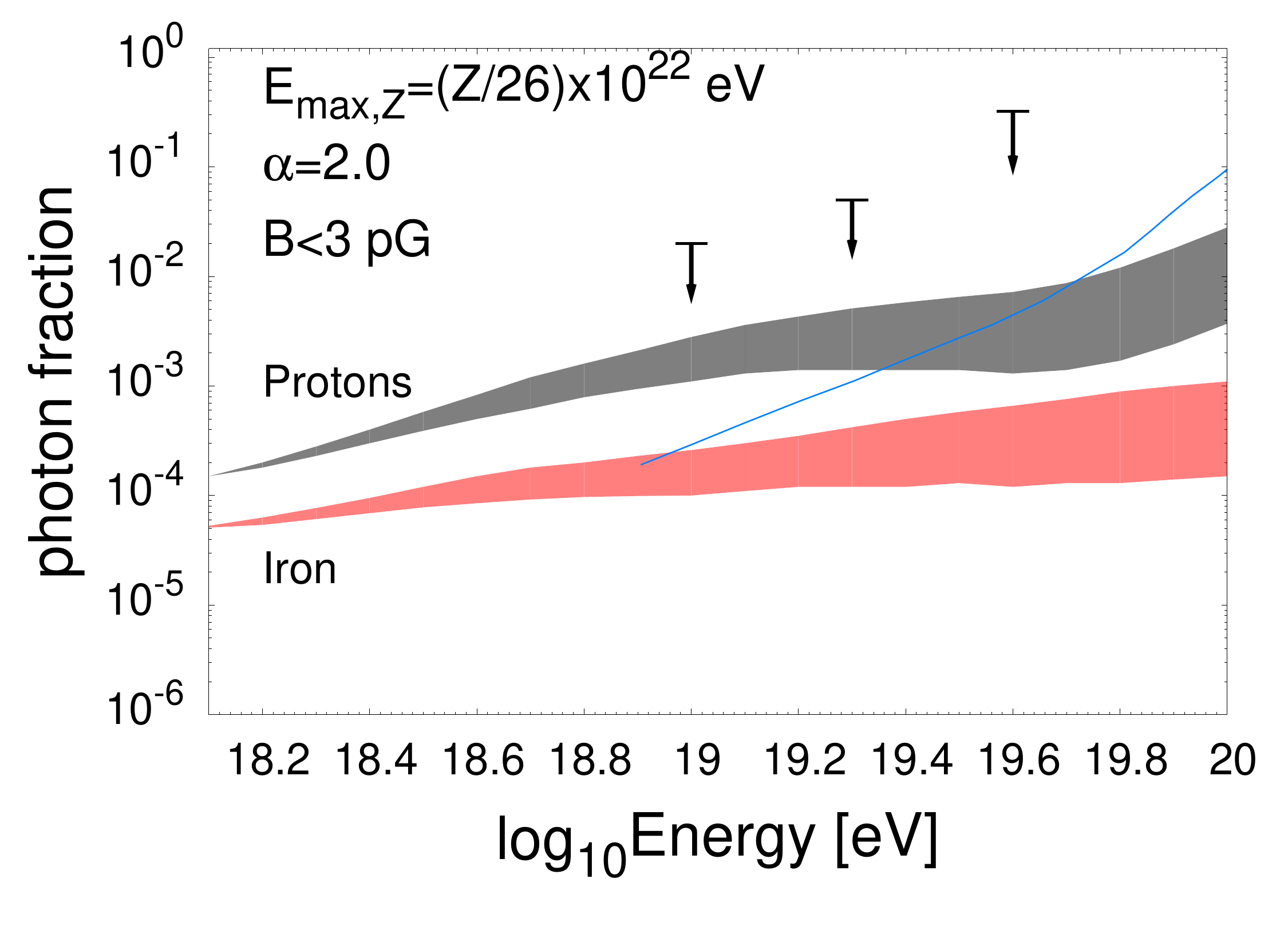}
\vspace{-0.3cm}
\caption{The fraction of ultra-high energy cosmic rays that are photons as a function of energy for the case of weak extragalactic magnetic fields ($<3 \times 10^{-12}$~G).
  Results are shown for two choices of the maximum injected
  energy and for models in which the cosmic ray sources inject
  uniquely protons, nitrogen, silicon, or iron nuclei. The bands reflect the range of the extragalactic radio backgrounds considered. 
  Also shown are the upper limits on the photon
  fraction from the Pierre Auger
  Observatory~\cite{augersdphotonfractionlimit} and its ultimate
  projected reach (blue line)~\cite{Risse:2007sd}.}
\label{window}
\end{center}
\end{figure*}

We show in Fig.~\ref{window} the photon fraction of UHECRs at Earth in
different models of the primary composition for the case of weak extragalactic magnetic fields ($<3~$pG).  If the primaries are largely protons, then the UHE photon fraction 
at $10^{19}$ eV ranges from $\sim 10^{-4}$ for $E_{\rm max}=10^{21}$ eV to $\sim 10^{-3}$
for $E_{\rm max}=10^{22}$ eV. The bands shown in the figure
represent the variation resulting from the range of radio backgrounds
considered. For comparison, we show the upper limits on the photon fraction set 
by the PAO \cite{augersdphotonfractionlimit} as well as its projected reach
(after 20 years of observation) \cite{Risse:2007sd}. We see that
proton dominated UHECR will likely provide a detectable photon
fraction so long as $E_{\rm max}$ is not too close to the GZK cutoff.

The situation is very different if the UHECRs are mostly heavy or
intermediate mass nuclei. Generally speaking, this leads to
approximately an order of magnitude suppression of the photon
fraction. If, for example, the UHECR sources inject only iron nuclei (as
shown in the lower frames of Fig.~\ref{window}), the photon fraction
never exceeds $\sim 3\times 10^{-4}$, and is thus beyond the reach of
the PAO. For intermediate mass nuclei at source, the photon fraction
is less suppressed, but is still considerably lower than for the
all-proton case. Note that all of the models considered here are consistent with
the cascade limit on the GeV-TeV photon flux and with bounds on the
cosmogenic neutrino flux \cite{evolution}.

\begin{figure*}[!t]
\begin{center}
\includegraphics[angle=0,width=0.325\linewidth,type=pdf,ext=.pdf,read=.pdf]{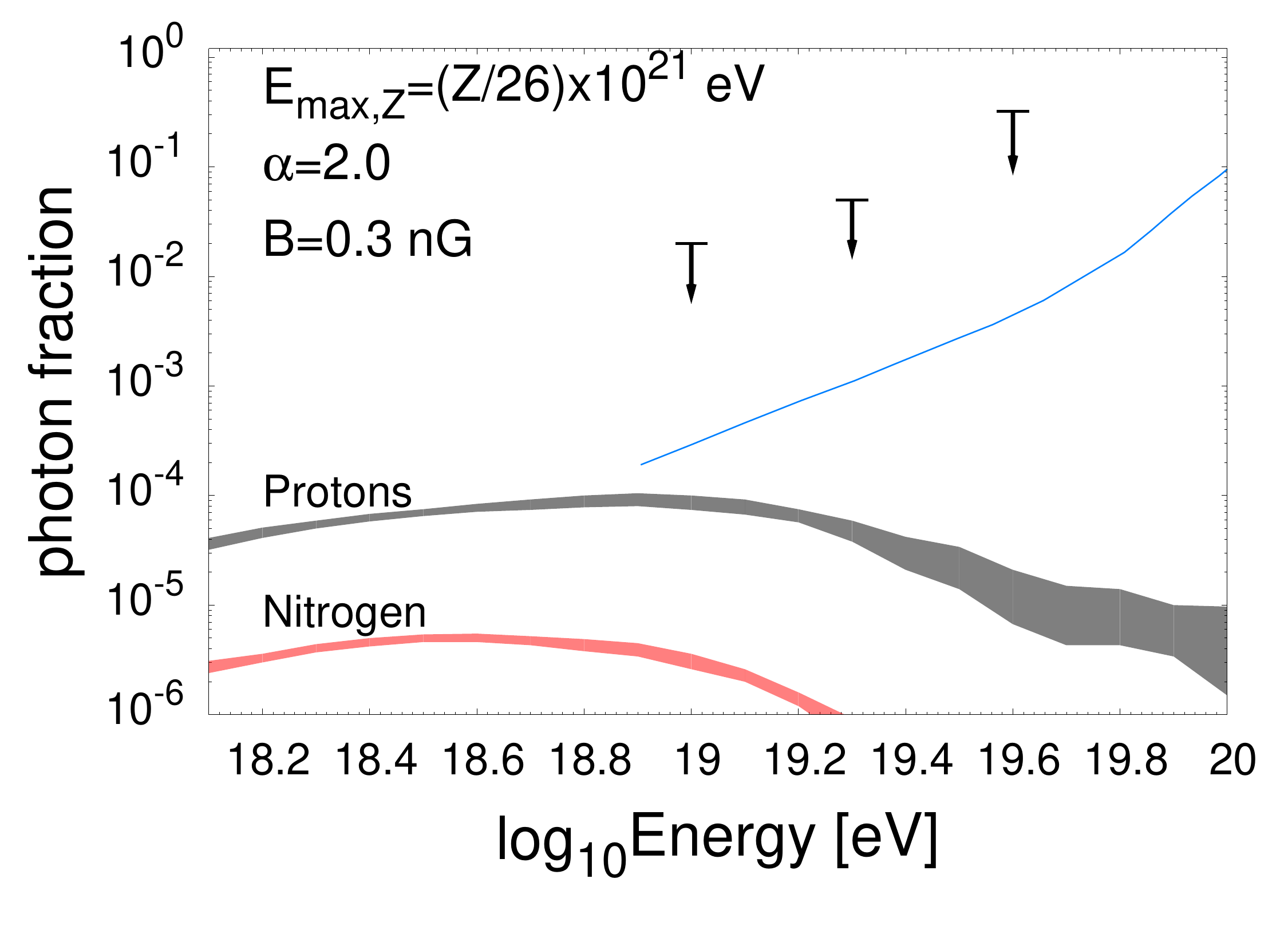}
\includegraphics[angle=0,width=0.325\linewidth,type=pdf,ext=.pdf,read=.pdf]{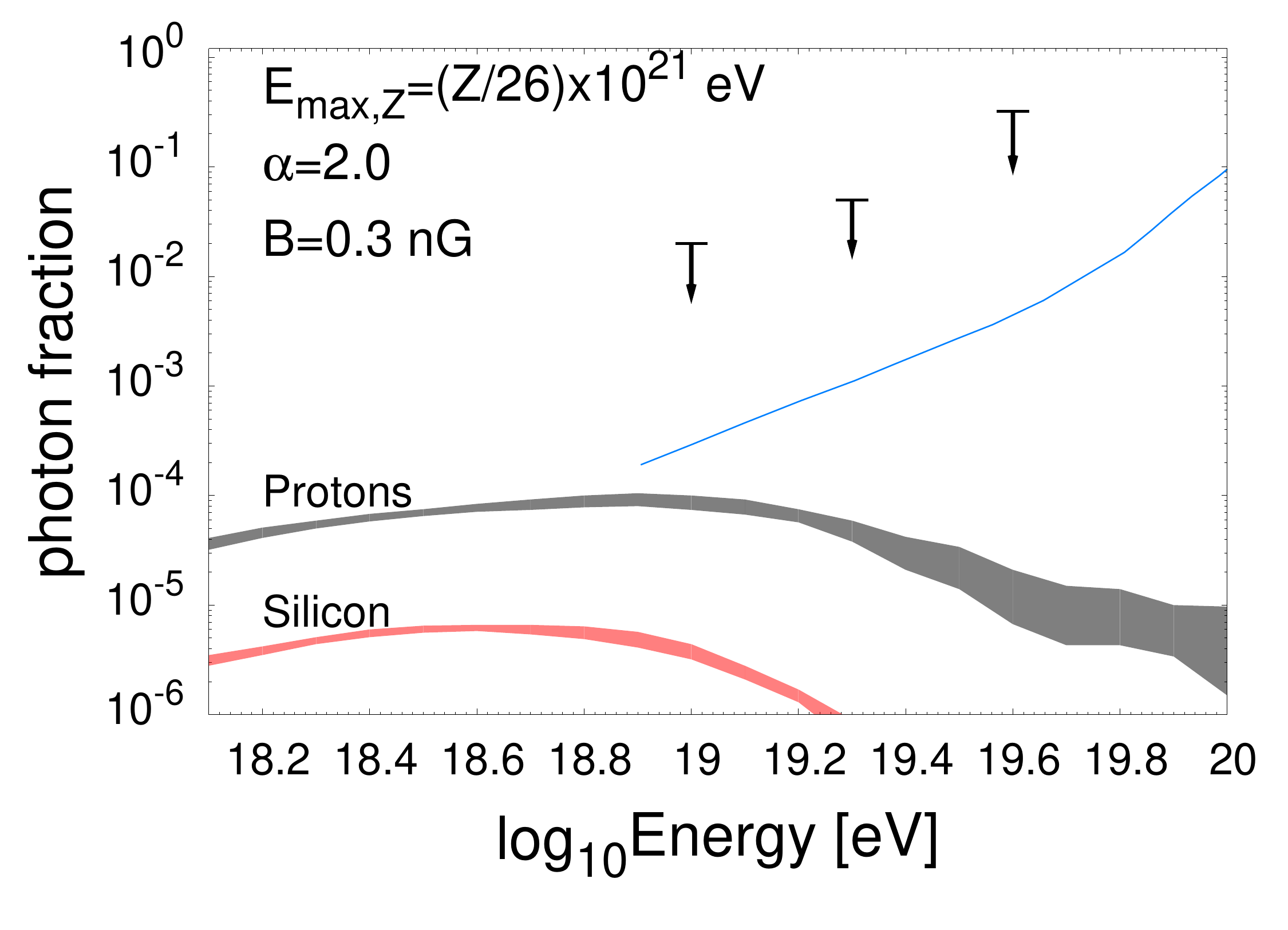}
\includegraphics[angle=0,width=0.325\linewidth,type=pdf,ext=.pdf,read=.pdf]{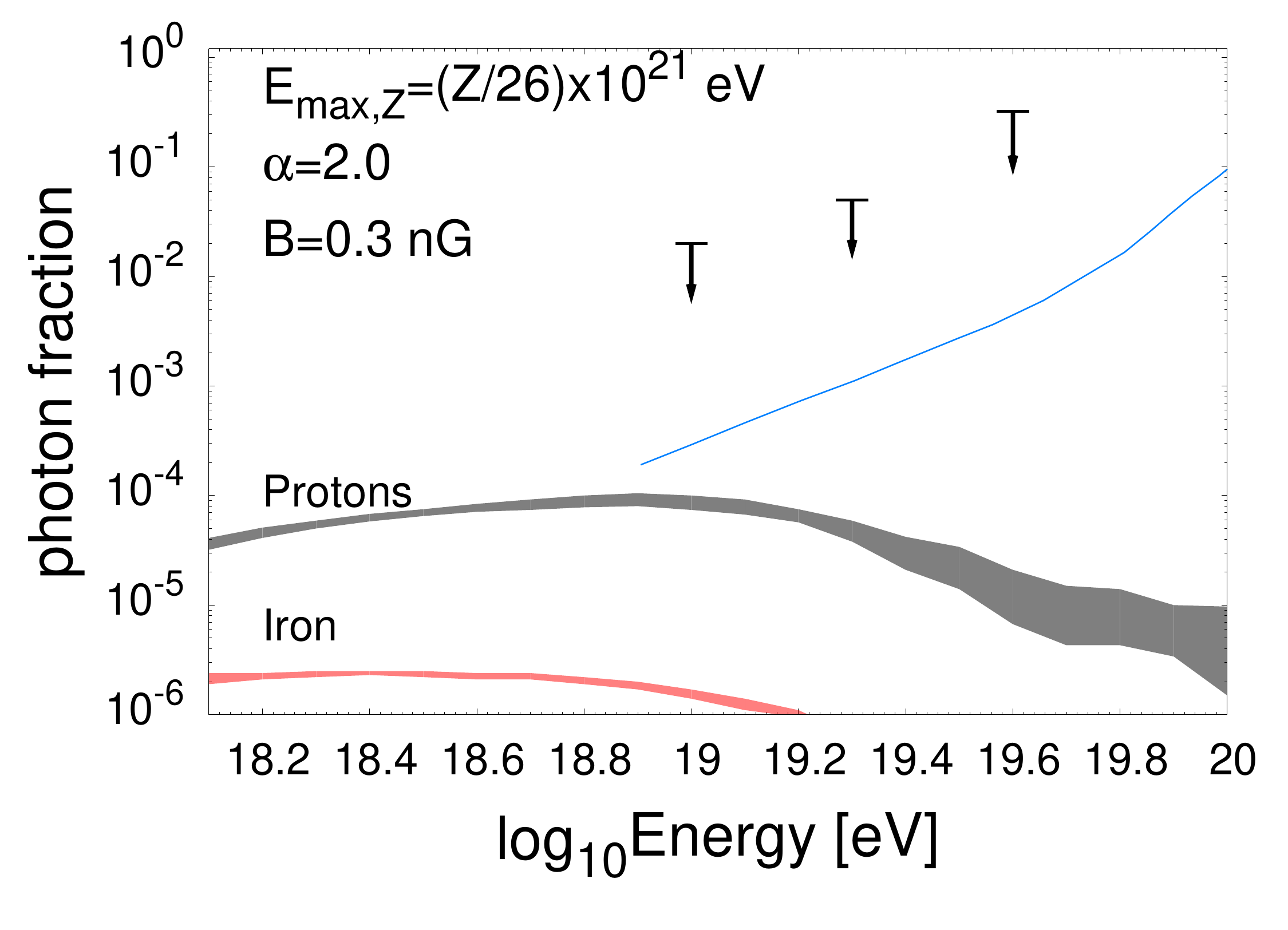}
\includegraphics[angle=0,width=0.325\linewidth,type=pdf,ext=.pdf,read=.pdf]{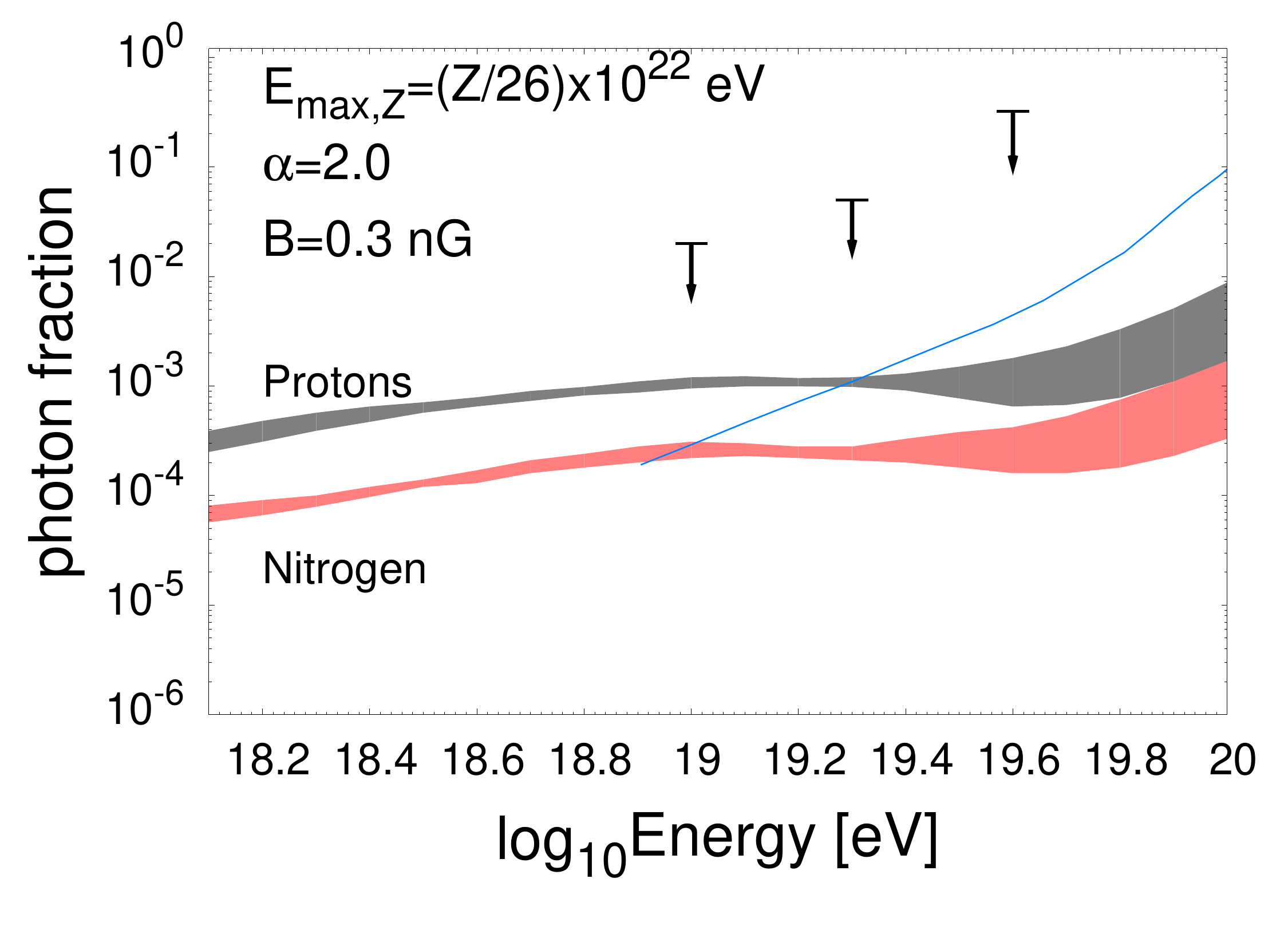}
\includegraphics[angle=0,width=0.325\linewidth,type=pdf,ext=.pdf,read=.pdf]{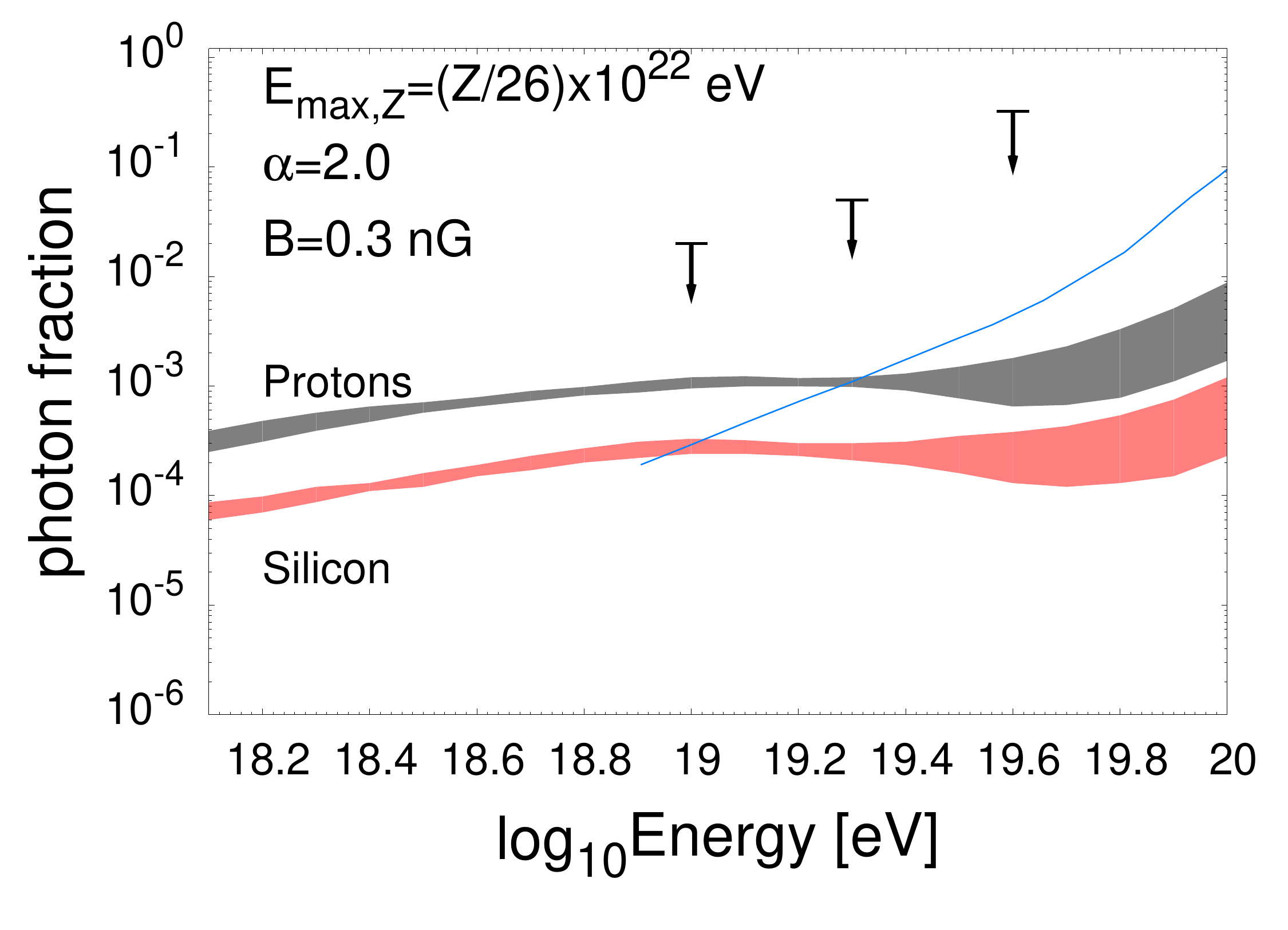}
\includegraphics[angle=0,width=0.325\linewidth,type=pdf,ext=.pdf,read=.pdf]{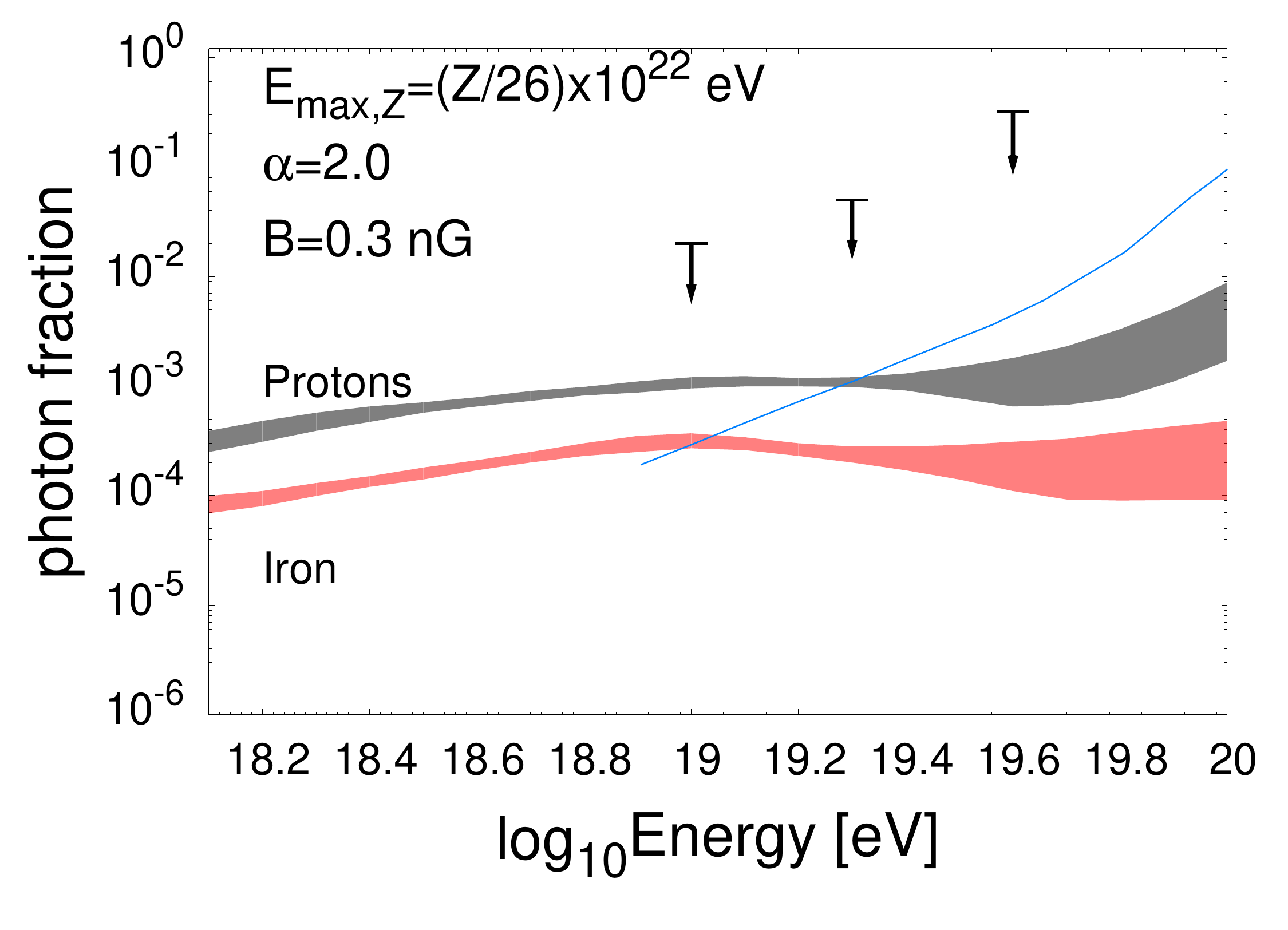}
\vspace{-0.3cm}
\caption{The same as in Fig.~\ref{window}, but for the case of 0.3~nG extragalactic magnetic fields.}
\label{window2}
\end{center}
\end{figure*}

In Fig.~\ref{window2}, we show the photon fraction of UHECRs at Earth in
different models of the primary composition for the case of 0.3~nG 
extragalactic magnetic fields. The effect of the presence of such a strong
extragalactic magnetic field is to increase the photon fraction
at energies near $\sim 10^{18}$~eV, and decrease it above
$10^{19}$~eV, as was previously suggested in Ref.~\cite{Taylor:2008jz}.

We note that our results differ somewhat from those previously presented in
Ref.~\cite{prior}. Whereas we find approximate agreement with Ref.~\cite{prior} for the cases of protons or iron nuclei, we disagree in the case of helium. In particular, we obtain a photon fraction in the case of helium nuclei that is between the values found in the proton and iron cases, whereas Ref.~\cite{prior} quotes
values below those found for iron nuclei. This is puzzling as the
photon fraction should predominantly depend upon the fraction of
fragmented protons produced locally ({\em i.e.} within
$\sim 100$~Mpc) with energies above the threshold for pion
production. Since a rigidity-dependent cutoff leads to a maximum fragmented proton energy proportional to $Z/A$, the photon fraction for heavier
nuclei should decrease monotonically with increasing $A$. Furthermore,
pair production losses further reduce the contribution from heavy nuclei relative
to lighter nuclei, and should thus decrease the photon fraction below
that for lighter nuclei. It appears that although the
authors of Ref.~\cite{prior} did consider photopion production by secondary
nucleons, they neglected pair production by protons and nuclei and
photopion production by secondary nuclei \cite{gelmini}.

{\it Summary}: We find that if ultra-high cosmic rays consist largely
of heavy or intermediate mass nuclei, then the cosmogenic photon flux will
be suppressed by about a factor of 10 relative to that expected for
proton primaries. This provides a means of potentially discriminating
between composition scenarios that is not subject to the
uncertainties associated with hadronic interaction models. As the
Pierre Auger Observatory continues to collect data, it is projected to
reach the sensitivity required to use this distinction to constrain
the chemical composition of the UHECRs. This would be complementary to
the information potentially provided by future measurements of the
cosmogenic neutrino flux which depends significantly on the
cosmological evolution of UHECR sources -- greater or fewer sources at
high redshifts would lead to a higher or lower neutrino flux,
respectively~\cite{evolution}. In contrast, since any observed ultra-high energy photons
must have originated within $\sim$100 Mpc, cosmological source
evolution cannot affect their flux.

{\it Acknowledgements:} We are grateful to Graciela Gelmini for
clarifications of earlier work and to Markus Risse for helpful
correspondence. DH is supported by the US Department of Energy,
including grant DE-FG02-95ER40896, and by NASA grant NAG5-10842. SS
acknowledges support by the EU Marie Curie Network ``UniverseNet''
(HPRN-CT-2006-035863).

\end{document}